\documentclass{article}
\usepackage{graphicx}
\usepackage[sort]{cite}
\usepackage{amssymb,latexsym,amsmath}
\begin{document}
\normalsize \twocolumn \columnsep=0.5cm
\title{Ion-Phosphate Mode in the DNA Low-Frequency Spectra}
\author{S. M. Perepelytsya$^{*}$, S. N. Volkov$^{**}$\\
 $^{*}$Kiev Shevchenco University, Department of Physics \\
 64 Volodymyrska Str.,Kiev 01033, Ukraine\\
 perepelytsya@univ.kiev.ua\\
$^{**}$Bogolyubov Institute for Theoretical Physics NASU,\\
 14-Á Metrologichna Str., Kiev 03143, Ukraine\\}
 \maketitle
 \begin{abstract}
 The vibrational dynamics of a DNA
molecule with counterions neutralizing the charged phosphate
groups have been studied. With the help of elaborated model the
conformational vibrations of the DNA double helix with alkaline
metal ions have been described both qualitatively and
quantitatively. For the complexes of DNA with counterions
$Li^{+}$, $Na^{+}$, $K^{+}$, $Rb^{+}$ and $Cs^{+}$ the normal
modes have been found, and a mode characterized by the most
notable ion displacements with respect to the DNA backbone has
been determined. The frequency of counterion vibrations has been
established to decrease as the ion mass increases. The results of
theoretical calculation have been showed to be in good agreement
with the experimental data of Raman spectroscopy.

Key words: DNA,ion, vibration, low-frequency spectra.

\end{abstract}
\textheight=22cm%\noindent
\section{Introduction}\label{S:Intro}
Starting with the early research by Watson and Crick
~\cite{Wotson} it has been a well known fact that a DNA
macromolecule in the natural form is a salt, a complex of nucleic
acid and metal cations. The ion environment of the double helix
plays the dominant role in the helical structure formation. The
concentration of the counterions in DNA solution determines the
macromolecule helix twisting, bending and the recognition of the
DNA sites by proteins and drugs
~\cite{Williams,Zigel,Blagoy,Zenger}. Therefore, metal cations are
of paramount importance in the processes of DNA functioning in
living systems.

Metal ions interacting with DNA may be divided in two groups
~\cite{Blagoy,Zenger}. The first group represents the ions
belonging to the diffusion atmosphere of the macromolecule.
Another group is the ions that are directly bound with different
structural elements of the double helix. The interaction of the
ions of this group is very specific and depends on their type.
Transition and alkaline-earth metal ions are bound principally
with nucleic base atoms, and the cations of alkaline metals being
bound with phosphate groups of the helix backbone ~\cite{Zenger}.

Under ordinary conditions the thermal fluctuations cause
vibrations of DNA structural elements and counterions around their
equilibrium positions. The vibrations of alkaline metal ions with
respect to the DNA backbone phosphates (ion-phosphate modes) are
likely to be visual and vastly intensive in vibrational spectra
because of homogeneity of backbone phosphate groups and interacted
cations. Determination of the character of these vibrations is of
great interest for understanding the counterion role in the DNA
helix mobility and conformational transformations.

The vibrations of the double helix structure elements may be
observed in the low-frequency region because of the massiveness of
DNA atomic groups and a relatively weak interaction between them.
In fact, the DNA vibration spectra show the set of modes in the
region lower then 250 $cm^{-1}$ \cite{Tominaga, Tominaga2, Urabe,
Urabe2, Weidlich1, Weidlich2, Weidlich3, Wittlin,Powell,Lindsay,
Lamba,Demarco}. The low-frequency spectra of DNA may be
conditionally divided in three ranges. The lowest range (10 -- 30
$cm^{-1}$) characterizes vibrations of the double helix backbone
~\cite{VolkovM,Weidlich1}. The DNA frequencies in this range
depend on conterion type ~\cite{Weidlich3,Weidlich1, Tominaga2},
and humidity ~\cite{Demarco,Weidlich1}. The middle range (60 - 120
$cm^{-1}$) is generated by vibrations of the hydrogen bonds and
intranucleoside mobility ~\cite{VolkovM,Weidlich1}.  A mode
strongly depending on counterion type has been observed in the
range within 120 -- 250 $cm^{-1}$ ~\cite{Weidlich2}. In this
frequency range an internal vibrational mode of the nucleic bases
can be also observed ~\cite{Lamba, Shishkin}. Therefore, it is
important to estimate a possible range of the DNA ion-phosphate
modes and their dependence on ion type.

To describe the low-frequency vibrational spectra of DNA
macromolecule the theoretical approach has been proposed
~\cite{VolkovM,Volkov}. However, in this approach the counterion
vibrations have not been taken into consideration. The problem of
ion effect on the DNA low-frequency modes has been studied in
works ~\cite{Powell, Prohofsky, Garcia}. The results have showed
that the DNA low-frequency vibrational spectra depend on
counterions, but the specific ion mode has not been found.

The purpose of this paper is to describe the low-frequency
vibrations of DNA with counterions, as well as to determine a
place of ion-phosphate mode and its dependence on counterion type.

\section{Model of DNA with counterions}\label{S:Model}

In frame of the approach ~\cite{VolkovM,Volkov} the improved model
of the DNA vibrations is elaborated ~\cite{Perepelytsya}. The
model presents DNA as a double chain of the backbone masses
$m_{0}$ ($PO_{2}+2O+C_{5}$) with connected pendulum-nucleosides
$m$ (sugar+base). The counterions are modelled as a masses
$(m_{a})$ bonded to the backbone. The nucleoside bases of
different chains connected by hydrogen bonds (Fig.1).
\begin{figure}[b!]
\small
\includegraphics[width=8cm,height=11cm]{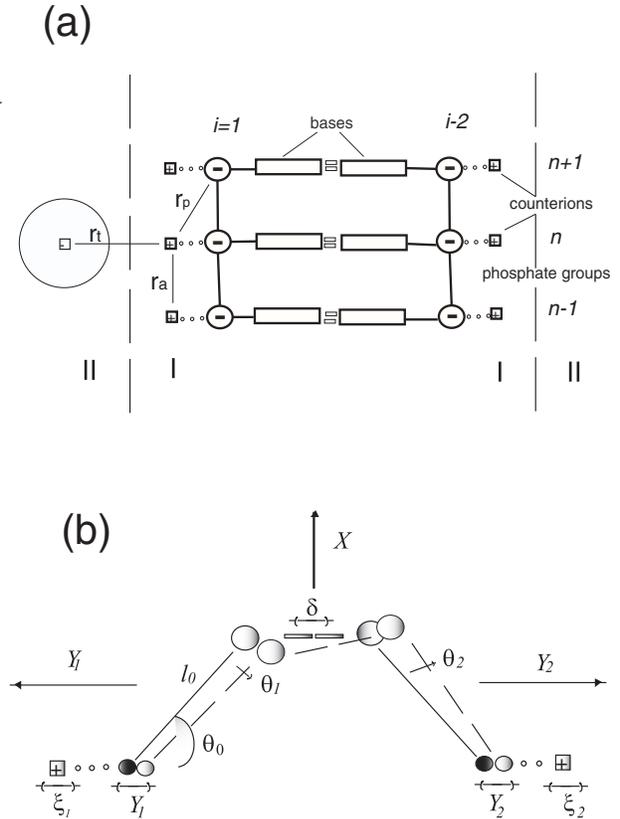}
\caption {The four mass model with couterion. (a) The monomer
links of the DNA. I -- the counterions that directly bind to the
DNA structure elements, II -- the diffusion atmosphere of the
macromolecule  (b) The displacements of the atomic groups and ions
in monomer link}\label{Fi:1}
\end{figure}
The geterogeneity of the DNA nucleosides is important for H-bond
stretching vibrations in  base pairs and for intranucleoside
mobility ~\cite{VolkovM, Volkov}. In present paper we intend to
find the place of the ion mode in the haul DNA low-frequency
spectra, and on this stage the geterogeneity of nucleosides and
the intranucleoside mobility are not considered.

Let us write the energy of the system:
\begin{equation}\label{E:energy}
  E=\sum_{n}\sum_{i=1}^2\left(K_{ni}+U_{ni}\right)+\sum_{n,n-1}U_{n,n-1},
\end{equation}
where $K_{ni}$, and $U_{ni}$ are the kinetic and the potential
energy of $n$-th monomer link of the system, and $U_{n,n-1}$ is
the interaction along the chain.

The expression for kinetic energy of the $n$-th monomer can be
written in the following way:
\begin{equation}\label{E:kin}
  K_{ni}=\frac{1}{2}M\dot{Y}_{ni}^2+\frac{1}{2}I\dot{\theta}_{ni}^2
+ml_{s}\dot{\theta}_{ni}\dot{Y}_{ni}
+\frac{1}{2}m_{a}(2\dot{Y}_{ni}\dot{\xi}_{ni}+\dot{\xi}_{ni}^2),
\end{equation}
where $M=m_{0}+m+m_{a}$; $Y_{n}$ -- the displacement of $n$-th
nucleoside and phosphate group mass; $I=ml^{2}+m_{a}r^{2}_{0}$ --
the inertia moment of nucleotide with counterion; $l$ is the
reduced length of the pendulum-nucleoside, and $r_{0}$ is the
equilibrium distance between ion and phosphate group; $\theta_{n}$
-- the angle of displacement of $n$-th nucleoside from equilibrium
state $\theta_{eq}$; $l_{s}=l\sin\theta_{eq}$; $\xi_{n}$ -- the
displacement of $n$-th ion from the equilibrium state.

For small displacements the potential energy of the monomer link
may be presented as a sum of three terms describing the energy of
H-bond stretching in the pair, the energy of torsion motions of
nucleoside around the backbone chains, and the energy of
ion-phosphate bond:
\begin{equation}\label{E:pot}
  U_{ni}=\frac{1}{2}\left(\alpha\delta_{n}^2+\beta\theta_{ni}^2+\gamma\xi_{ni}^2\right).
\end{equation}
Here $\alpha$ is the force constant of hydrogen bond stretching in
pairs,
$\delta_{n}\thickapprox{l_{s}(\theta_{1n}+\theta_{2n})+Y_{1n}+Y_{2n}}$
is the base pair stretching,  $\beta$ -- is the force constant of
nucleoside vibration relatively phosphate group, $\gamma$ is the
constant of ion-phosphate vibration.

In frame of the DNA conformational vibrations model (Fig. 1) we
are looking for the normal modes of the lattice of double helix
with counterions.  To compare our calculations with experimental
data let as consider a limited long-range vibrations of optic type
that have to be seen in a spectra of real objects. In this case,
it is enough to consider the homogeneous displacements of the
system monomer links, where the interaction between monomers along
the chain ($U_{n,n-1}$ ) may be waved.

Using the expressions for the energy ~(\ref{E:energy}),
~(\ref{E:kin}), and ~(\ref{E:pot}) let us write the Lagrange
equations of motion for this system. For the conveniens of the
following consideration we introduce the variables:
\begin{eqnarray}\label{E:Coordinates}
Y=Y_{1}+Y_{2},\mbox{    }y=Y_{1}-Y_{2},\\\notag
\theta=\theta_{1}+\theta_{2},\mbox{}\eta=\theta_{1}-\theta_{2},\\
\xi=\xi_{1}+\xi_{2},\mbox{    }\rho=\xi_{1}-\xi_{2}\notag.
\end{eqnarray}
In variables ~(\ref{E:Coordinates}) the equations of motion for
whole system brake up into two systems of connected vibrations:
\begin{equation}
 \ddot{Y}+\frac{ml_{s}}{M}\ddot{\theta}+\frac{m_{a}}{M}\ddot{\xi}=
-\alpha_{0}(l_{s}\theta+Y);\notag
\end{equation}
\begin{equation}\label{E:System1}
 \ddot{\theta}+\frac{ml_{s}}{I}\ddot{Y}
=-\beta_{0}\theta-\alpha_{0}\frac{Ml_{s}}{I}(l_{s}\theta+Y);
\end{equation}
\begin{equation}
 \ddot{Y}+\ddot{\xi}=-\gamma_{0}\xi,\notag
\end{equation}

\begin{equation}
 \ddot{y}+\frac{ml_{s}}{M}\ddot{\eta}+\frac{m_{a}}{M}\ddot{\rho}=0;\notag
\end{equation}
\begin{equation}\label{E:System2}
 \ddot{\eta}+\frac{ml_{s}}{I}\ddot{y}=-\beta_{0}\eta;
\end{equation}
\begin{equation}
 \ddot{y}+\ddot{\rho}=-\gamma_{0}\rho.\notag
\end{equation}
Here is $\alpha_{0}=2\alpha/M$, $\beta_{0}=\beta/I$,
$\gamma_{0}=\gamma/m_{a}$.

The normal modes of the system ~(\ref{E:System1}) and
~(\ref{E:System2}) let us find in form: $q=q_{0}e^{-i\omega t}$.
From the system of equations ~(\ref{E:System1}) the equation for
frequencies is obtained:
 \small
\begin{equation}\label{E:w1}
 (\gamma_{0}-\omega^{2})\left[(\alpha_{0}-\omega^{2})(p_{0}-\omega^{2})- \frac{Ml_{s}^{2}}{I}(\alpha_{0}-
 \omega^{2}m/M)^{2}\right]-
\end{equation}
\begin{equation}
 -\frac{m_{a}}{M}\omega^{4}(p_{0}-\omega^{2})=0.\notag
\end{equation}
\normalsize
 Here $p_{0}=\alpha_{0}Ml_{s}^{2}/I+\beta_{0}$. The
solutions of equation ~(\ref{E:w1}) are the modes $\omega_{1}$,
$\omega_{2}$, and $\omega_{3}$.

From ~(\ref{E:System2}) the following equation is obtained:
\begin{equation}\label{E:w2}
\omega^{4}\mu+\omega^{2}(\beta_{0}^{\prime}+\gamma_{0}^{\prime})+\gamma_{0}\beta_{0}=0
\end{equation}
Here $\beta^{\prime}_{0}=\beta_{0}(m_{a}/M-1)$,
$\gamma^{\prime}_{0}=\gamma_{0}(m^{2}l^{2}_{s}/MI-1)$,
$\mu=1-m_{a}/M-m^{2}l^{2}_{s}/MI$. The solutions of ~(\ref{E:w2})
are the modes $\omega_{4}$, and $\omega_{5}$:
\begin{equation}\label{E:Omega45}
\omega^{2}_{4,5}=\frac{-(\beta^{\prime}_{0}+\gamma^{\prime}_{0})\pm\sqrt{(\beta^{\prime}_{0}+\gamma^{\prime}_{0})^{2}-4\mu\beta_{0}\gamma_{0}}}
{2\mu}.
\end{equation}

Let us analyze the modes $\omega_{4,5}$. Taking into account that
for all counterion type $m_{a}/M$ and $m^{2}l^{2}_{s}/MI$ are
rather small (about 0.1), formula ~(\ref{E:Omega45}) may be
transformed:
\begin{equation}\label{E:Omega4}
\omega^{2}_{4}\approx\gamma_{0},
\end{equation}
\begin{equation}\label{E:Omega5}
\omega^{2}_{5}\approx\beta_{0}.
\end{equation}
Expressions ~(\ref{E:Omega4}) and ~(\ref{E:Omega5}) show that the
modes $\omega_{4}$ and $\omega_{5}$ depend, respectively, on ion
vibration constant, and constant of backbone vibration. Analyzing
the ratio between amplitudes obtained from ~(\ref{E:System2}):
\begin{equation}\label{E:Amplitude1}
\frac{y_{0}}{\eta_{0}}=\frac{(\beta_{0}-\omega^{2})I}{\omega^{2}ml_{s}},
\end{equation}
\begin{equation}\label{E:Amplitude2}
\frac{\rho_{0}}{\eta_{0}}=\frac{(\beta_{0}-\omega^{2})I}{(\gamma_{0}-\omega^{2})ml_{s}},
\end{equation}
it is clear that on the mode $\omega_{4}$ the amplitude of ion
vibration ($\rho_{0}$) is much lager in comparison with the
amplitudes of phosphate ($y_{0}$) and nucleoside ($\eta_{0}$)
motions. So the mode $\omega_{4}$ characterizes mostly by ion
displacements. Therefore we consider that  this mode is the mode
of ion-phosphate vibrations.

Let us analyze the equation ~(\ref{E:w1}). As seen, the form of
equation ~(\ref{E:w1}) depends on the value of counterion mass
($m_{a}$). Taking into account that for light ions ($Li^{+}$ and
$Na^{+}$) the ratio $m_{a}/M$ is very small (about 0.05), the last
term in equation ~(\ref{E:w1}) may be waved:
\begin{equation}\label{E:w1approx}
 (\gamma_{0}-\omega^{2})\left[(\alpha_{0}-\omega^{2})(p_{0}-\omega^{2})-
 \frac{Ml_{s}^{2}}{I}(\alpha_{0}-\omega^{2}m/M)^{2}\right]\approx0.
\end{equation}
 One of the solution of ~(\ref{E:w1approx}) is
$\omega_{1}\approx\sqrt{\gamma_{0}}$. The amplitude of ion
vibration on the mode $\omega_{1}$ is the largest  in comparison
with the amplitudes of phosphate ($Y_{0}$) and nucleoside
($\theta_{0}$) motions.  So we have found that for light
counterions the mode $\omega_{1}$ may be also considered as a mode
of ion vibration. It means that for light counterions in the DNA
low-frequency spectra there is a twice degenerated mode of
ion-phosphate vibrations
($\omega_{1}=\omega_{4}\approx\sqrt{\gamma_{0}}$).

The analysis for heavy ions ($Rb^{+}$ and $Cs^{+}$) chows that on
the mode $\omega_{1}$ the amplitude of H-bond stretching is the
largest. So the structure of the DNA low-frequency spectra is
qualitatively different for light and heavy counterions.

For further analysis of the ion mode let us determine the constant
of ion vibration. Taking into account that counterions form the
regular structure around the sugar-phosphate backbone of the DNA
such system may be considered as the ion crystal. According to the
classical study, the energy of ion interaction in crystal can be
presented by potential:
\begin{equation}\label{E:crystal}
  V(r)=-\frac{M_{\alpha}e^2}{4\pi\varepsilon\varepsilon_{0}r}+B\exp{\left(-\frac{r}{b}\right)}
\end{equation}
where $M_{\alpha}$ is a Madelung constant, $e$ is the ion charge,
$\varepsilon$ is the dielectric constant, $r$ is the distance
between charges, $B$ and $b$ characterize the repulsion between
ions. The first term describes the electrostatic attraction and
the second -- Born-Mayer repulsion. Expanding the energy
~(\ref{E:crystal}) in the row on small displacements from their
equilibrium positions $\xi=r-r_{0}$, $V\approx\gamma\xi^{2}/2$ the
constant of ion vibrations is obtained:
\begin{equation}\label{E:gama}
\gamma=\frac{M_{\alpha}e^2}
{4\pi\varepsilon\varepsilon_{0}r_{0}^3}\left(\frac{r_{0}}{b}-2\right).
\end{equation}
Using the expression for the constant of ion vibration
~(\ref{E:gama}) and ~(\ref{E:Omega4}) the approximate formula for
the mode of ion vibration is obtained:
\begin{equation}\label{E:IonMode}
\omega_{ion}\approx\sqrt{\frac{M_{\alpha}e^2}
{4\pi\varepsilon\varepsilon_{0}m_{a}r_{0}^3}\left(\frac{r_{0}}{b}-2\right)}.
\end{equation}
Analyzing ~(\ref{E:IonMode}) we can see that the ion-phosphate
mode strongly depends of the equilibrium distance $r_{0}$, and ion
mass $m_{a}$. Thus we can make the conclusion that the ion mode
has to decrease with increasing of the ion radius:
\begin{equation}\label{E:IonModeLine}
\omega_{Li}>\omega_{Na}>\omega_{K}>\omega_{Rb}>\omega_{Cs}.
\end{equation}
The ion-phosphate mode is also depended on the Madelung constant
$M_{\alpha}$. This constant is the lowest for dipole
($M_{\alpha}=1$) and is big enough for the $NaCl$ ion crystal
($M_{\alpha}=1.748$). In DNA the value of the Madelung constant is
not known, but we can say that the Madelung constant is depended
on the ion concentration, and in dilute solution it will be
another then in the saturated solution. It means that the
ion-phosphate mode has to be depended of the ion force of the
solution.

The constant of ion vibration also depends on the other parameters
($b$, $\varepsilon$) that will be considered in the following
section.

\section{Constant of ion vibration}\label{S:IonBond}
In this paper we study the conformational vibrations of DNA with
alkaline metals. So the equilibrium distance $r_{0}$ is estimated
as a sum of Pauling radiuses of ion and oxygen atom of the
phosphate group (Table 1). The parameter $b$ is taken equaled to
values that are known for ion crystals $LiF$, $NaF$, $KF$, $RbF$,
$CsF$ ~\cite{Melvin} (Table 1).

\begin{center}
\small
Table 1. The parameters for ion vibration constant
\begin{tabular}{c|ccccc}\hline

                 Parameter&$Li^{+}$&$Na^{+}$&$K^{+}$&$Rb^{+}$&$Cs^{+}$\\\hline
$r_{0}$($\AA$)&2.00&2.35&2.73&2.88&3.01\\
$b$ ($\AA$)&0.329&0.316&0.327&0.329&0.313\\
$\varepsilon$~\cite{Hingerty}&5.0&6.4&8.2&9.0&10.1\\
$\varepsilon$~\cite{Lavery}&1.3&1.5&1.8&1.9&2.1\\\hline
$M_{\alpha}$ ($\varepsilon=const.$)&1.307&1.360&1.417&1.439&1.470\\
$M_{\alpha}$ ($\varepsilon$ ~\cite{Hingerty})&1.045&1.067&1.099&1.114&1.136\\
$M_{\alpha}$ ($\varepsilon$
~\cite{Lavery})&1.050&1.066&1.089&1.099&1.116\\\hline
  \end{tabular}
\end{center}

Let us estimate the dielectric constant. The value of this
constant depends on the distance between ions and the nature of
environment.  The dielectric constant determination for the DNA
presents some difficulties ~\cite{Hingerty, Lavery, Mazur,Wang,
Olson,Ramachandran}. The value of vacuum dielectric constant
($\varepsilon=1$) is not suitable for description of electrostatic
interaction in DNA because in this case the potential near DNA is
too high ~\cite{Mazur}. Very often for solving the
Puasson-Boltzman equeation the dielectric constant of the
DNA-solution system is considered between 2 and 4 ~\cite{Mazur}.
Hingerty with coworkers developed an approach where the dielectric
constant presents as a function of a distance ~\cite{Hingerty}:
\begin{equation}\label{E:Hingerty}
  \varepsilon(r)=78-77(r/2.5)^{2}\exp(r/2.5)(\exp(r/2.5)-1)^{-2}.
\end{equation}
Lavery with coworkers were improved the function of Hingerty
\cite{Lavery}:
\begin{equation}\label{E:Lavery}
 \varepsilon(r)=\varepsilon_{\infty}-\frac{\varepsilon_{\infty}-1}{2}[(sr)^{2}+2sr+2]e^{-sr},
\end{equation}
were $\varepsilon_{\infty}=78$, and $s=0.16$. There are also exist
other distant dependent dielectric functions ~\cite{Mazur,Olson,
Wang}, but functions of Hingrty ~(\ref{E:Hingerty}) and Lavery
~(\ref{E:Lavery}) are most common used. The dielectric constants
calculated by formulas ~(\ref{E:Hingerty}) and ~(\ref{E:Lavery})
 are shown in Table 1.

Let us estimate the Madelung constant $M_{\alpha}$. This constant
characterizes the interaction of one ion with the other charges of
the system. In our model the charges of the phosphate groups with
counterions that neutralize them are considered. Also it is taken
into account that in the solution there exist negatively charged
ions of $Cl^{-}$. This anions are situated at the distance about
$7$ $\AA$ ~\cite{Blagoy}. As usual long-range electrostatic
interactions in such systems as DNA are catted off on distance
about $10$ $\AA$ ~\cite{Nilson} that is approximately equaled to
the distance between two neighbor phosphate groups of one polymer
chain. That is why only the charges of neighbor ion-phosphate
pairs and $Cl^{-}$ anions are considered in Madelung constant
calculations. The following formula for Madelung constant is
obtained:
\begin{equation}\label{Madelung}
M_{\alpha}=1+\frac{\epsilon (r_{0})r_{0}}{\epsilon
(r_{t})r_{t}}+2\frac{\epsilon (r_{0})r_{0}}{\epsilon
(r_{p})r_{p}}-2\frac{\epsilon (r_{0})r_{0}}{\epsilon (r_{a})r_{a}}
\end{equation}
Here $\epsilon (r)$ is the dielectric constant, $r_{t}$ is the
distance between cation and $Cl^{-}$ anion of the solution,
$r_{p}$ is the distance between cation and neighbor phosphate
group of the same polymer chain, and $r_{a}$ -- the distances
between neighbor cations of the same polymer chain (Fig.1). The
first term in ~(\ref{Madelung}) describes the interaction between
cation and $Cl^{-}$ anion of the solution, the second term
describes the interaction between cation and neighbor phosphate
group, and the third term describes the interaction between
neighbor cations. The values of the Madelung constants calculated
by formula ~(\ref{Madelung}) are shown in Table 1.

\section{Ion-phosphate mode}\label{S:IonMode}
 Using the parameters shown in Table 1 and formula
~(\ref{E:gama}) the values of the constant of ion vibrations
($\gamma$) is estimated. The frequencies of ion-phosphate
vibrations ($\omega_{4}$) calculated by expressions
~(\ref{E:Omega45}) are shown in Table 2.

\begin{center}
\small Table 2. The frequencies of ion-phosphate vibration for
\textit{B-}DNA ($cm^{-1}$)

\begin{tabular}{c|ccccc}\hline
                 Method &$Li^{+}$&$Na^{+}$&$K^{+}$&$Rb^{+}$&$Cs^{+}$\\\hline
$\varepsilon=2$ &444&233&162&112&93\\
$\varepsilon=4$&314&165&114&79&66\\
 $\varepsilon$ ~\cite{Lavery}&488 &237&151&100&80\\
$\varepsilon$ ~\cite{Hingerty}&252&115&70&46&37\\\hline
~\cite{Weidlich2}&237$^{*}$ &230$^{*}$ &150&110&95\\
 ~\cite{Lamba}&&235&&&\\\hline
  \end{tabular}
\end{center}
Note: The asterisk $^{*}$ marks the value determined by us from
the spectra of work ~\cite{Weidlich2} \vspace{\baselineskip}

Analyzing the results shown in Table 2 we can say that the best
agreement with the experimental data gives the value of
$\varepsilon=2$ and  Lavery dielectric function ~\cite{Lavery}. In
case of ion $Li^{+}$ the calculated frequency of vibration differ
sufficiently from the experimental data. The reason of this
situation may be connected with effects of ion hydratability. The
ion $Li^{+}$ has a very small weight  ($7$ $a. u. m.$) and radius
(0.6 $\AA$) in compare with molecules of water. Therefore,  the
influence of water on the lithium vibrations may be very large.
Our estimation $Li^{+}$ frequency when it is in complex with one
$(\omega_{4}^{Li}=267 cm^{-1})$, two $(209 cm^{-1})$, and three
$(180 cm^{-1})$ molecules of water show that the frequency value
became sufficiently lower. As seen if one consider that lithium
ion moves in the solution together with water molecules the
frequency become very close to the experimental value.

For understanding of the low-frequency spectra dependence on
counterion type the frequencies of all five branches of vibrations
$\omega_{1}$, $\omega_{2}$, $\omega_{3}$, $\omega_{4}$, and
$\omega_{5}$ are calculated (Fig. 2).
\begin{figure}[t!]
\includegraphics[width=8cm]{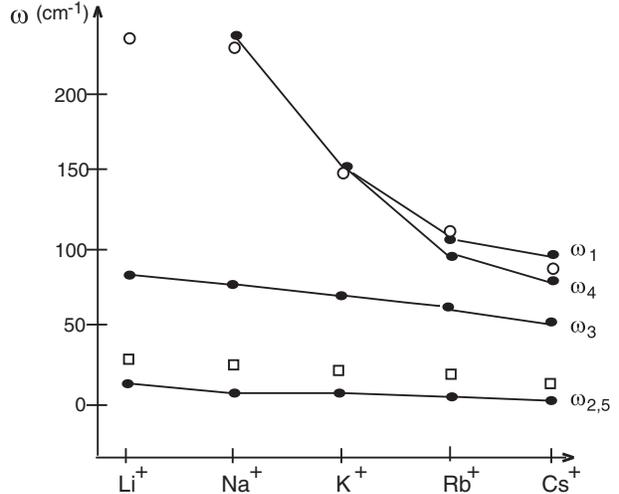}
 \caption{The
dependence of the low-frequency modes on counterion type.
$\bullet$ -- our results, \textbf{o} -- ~\cite{Weidlich2}, $\Box$
-- ~\cite{Weidlich3}.}\label{Fi:2}
\end{figure}
As we can see all low-frequency modes obtained with the help of
proposed approach decrease with increasing of counterion mass. The
ion mode ($\omega_{1}$, $\omega_{4}$) is the most sensitive to
counterion type. Our calculations show the splitting of the ion
mode for heavy ions  that was qualitatively obtained above. The
theoretical results for the lowest mode ($\omega_{2,5}$) and for
the ion mode ($\omega_{4}$) correlate with the experimental data.
The counterion dependence of the H-bond stretching mode
($\omega_{3}$) is not studied experimentally.

In the conclusion we can say that the specific mode of
ion-phosphate vibration was found. It is in frequency range 90 --
250 $cm^{-1}$ and decrease with increasing of ion radius. This
mode is very sensitive to the counterion type and to the ion force
of the solution. As the results presented in fig. 2 show all
low-frequency modes of the DNA macromolecule depend of counterion
type.

\pagebreak[4]

\end{document}